\documentclass[12pt,a4paper]{article}
\usepackage[pdftex,unicode]{hyperref}

\usepackage{mathtext}
\usepackage[T2A]{fontenc}
\usepackage[utf8]{inputenc}
\usepackage[english]{babel}

\usepackage{amsmath,amssymb}
\usepackage{amsxtra}
\usepackage{color}
\usepackage{array}
\usepackage{graphicx}

\def\eps{\varepsilon}
\def\ZZ{\mathbb{Z}}
\def\RR{\mathbb{R}}

\renewcommand{\exp}[1]{{\rm e}^{#1}}

\newcommand{\ii}{\mathfrak{i}}
\newcommand{\kap}{{\text{\it\ae}}}
\newcommand{\wigner}[3]{\left(
\begin{array}{ccc}#1&#2&#3\\0&0&0\end{array}
\right)}

\newcommand{\prob}{{\bf P}}
\newcommand{\me}{{\bf E}}
\newcommand{\var}{{\bf Var}}
\newcommand{\erf}{{\rm Erf}}
\newcommand{\heaviside}{\Theta}

\newcommand{\config}{\mathfrak{C}}
\newcommand{\statves}{\mathfrak{g}}

\newcommand{\ev}{\mbox{\,eV}}
\newcommand{\gccm}{\mbox{\,g$/$cm$^3$}}

\begin{document}

\title{A Detailed Configuration Accounting for the Hot Plasma Bound-Free Opacity Calculation}
\author{A.~V.~Nadezhkina$^1$, M.~A.~Vronskiy$^{1,2\,}$\footnote{Corresponding author. E-mail: mavronskiy@vniief.ru, mavronskii@mephi.ru}}
\date{
\parbox{12cm}{
{\small\it
$^1${Russian Federal Nuclear Center --- All-Russian Research Institute of\\Experimental Physics}\\
$^2${Sarov Physics and Technology Institute branch of the National Research\\Nuclear University MEPhI (Moscow Engineering Physics Institute). SarPhTI NRNU MEPhI}
}
}
}

\maketitle
\begin{abstract}

The calculation of the hot plasma bound--free opacity according to the average atom models often leads to a noticeable effect of initial configuration on the shell ionization threshold. 
For the related problem of taking into account the impact of configurations while calculating bound-bound opacity, G.~Hazak and Y.~Kurzweil developed a method within the Super Transition Arrays approach. Fourier transform method is the adaptation of their method for a Detailed Configurations Accounting.
The bound--free opacity coefficient qualitative behaviour makes it somewhat difficult to construct a direct analogue of Fourier transform method of Detailed Configurations Accounting for its calculation.
We use the probabilistic reformulation of the method to obtain the expressions for the bound--free opacity coefficient that take into account the spread of ionization thresholds due to the shell occupation numbers fluctuations.
\end{abstract}

\section{Introduction}
The spectral bound-free opacity of hot plasma describes photon absorption during transitions of electrons from bound to free states.
The bound-free opacity is an essential component of overall radiation opacity. For some frequency ranges its contribution may be predominant. 
For evaluating bound-free opacity coefficient, average-atom models are widely used. 
In these models, the positions of the opacity thresholds, i.e. the ionization energies of the electron shells, may depend significantly on the configuration, in which the transition occurs.
Therefore, to describe the behavior of bound-free opacity near thresholds, it is necessary to take into account the distribution of electron configurations.
This account is equivalent to the account of the fluctuations of the shells occupation numbers.
It is usually carried out either by replacing transitions from different configurations with transitions from an average configuration with an additional Gaussian spread of the threshold energy (see \cite{Dragalov1989}, \cite{Nikiforov2005}), or by combining configurations into superconfigurations that allow for enumeration, with the replacement of the ionization of the configuration with that of the superconfiguration (STA approach, see \cite{BarShalom1996}).

For the related problem of bound-bound opacity evaluation it is neccessary to take into account the impact of the electronic shell occupation numbers fluctuations on the transition energies. 
In the series of papers \cite{Hazak2012}, \cite{Kurzweil2013}, \cite{Kurzweil2016}, G.~Hazak and Y.~Kurzweil developed the method of Configurationally Resolved Super Transition Arrays (STA), that allows the refinement of that impact within STA approach. 
One of the features they make use of is the possibility to express the opacity coefficient as a Fourier transform. The direct application of Hazak --- Kurzweil method for the bound-free opacity coefficient is somewhat difficult: its qualitative behaviour is inappropriate to express it as a numerical Fourier transform. 
In \cite{Arapova2025}, the Hazak --- Kurzweil method adaptation (the Fourier transform method of the Detailed Configuration Accounting) was considered. For this method, the probabilistic representation for the bound-bound opacity coefficient was proposed. A similar representation with minor modifications is valid for the bound-free opacity.
To obtain expressions convenient for calculations, we replace the Fourier transform of the product of functions with a convolution. 
This replacement reduces the problem to finding the distribution function of a random variable from its characteristic function. 

\section{Assumptions}
Suppose we are given spherically symmetric potential $V(x)$ vanishing at infinity and the corresponding finite set of bound electron shells (energies and normalized two-component Dirac radial wavefunctions) indexed by natural numbers
$$
\eps_j,\ (f_j(x),g_j(x)),\ \int_0^{\infty}(f_j^2(x)+g_j^2(x))dx = 1,\ j=1,\ldots,M.
$$
Every index $j$ corresponds to Dirac quantum number $\kap_j\in\ZZ\setminus\{0\}$ and statistical weight $\statves_j = 2|\kap_j|$. 
Also, let $(f_{\eps\kap}(x),g_{\eps\kap}(x))$ be the one-electron Dirac radial wavefunctions corresponding to the free states with energy $\eps>0$ and Dirac quantum number $\kap$. Suppose that these wavefunctions are normalized over the energy scale
$$
\int_0^{\infty}(f_{\eps\kap}(x)f_{\eps'\kap}(x)+g_{\eps\kap}(x)g_{\eps'\kap}(x))dx = \delta(\eps-\eps')
$$
that is equivalent to 
\begin{equation}\label{ContinuousNormalize}
\pi\sqrt{\frac{\eps}{\frac{2me^4}{\hbar^2}+\alpha^2\eps}}\left(f_{\eps\kap}^2(x)+\frac{\frac{2me^4}{\hbar^2}+\alpha^2\eps}{\alpha^2\eps}g_{\eps\kap}^2(x)\right)\to 1\mbox{ when }x\to\infty,
\end{equation}
We exploit the usual notations $e$, $m$, $\hbar$, $c$, $a_0=\frac{\hbar^2}{me^2}$, $\alpha=\frac{e^2}{\hbar c}\approx 1/137$ for electron charge, electron mass, Planck's constant, speed of light, Bohr's radius and fine structure constant, respectively. 
Denote the set of all electronic configurations by 
$$
\config=\prod_{j=1}^{M}\{0,1,\ldots,\statves_j\}=\Bigl\{
{\bf n}=(n_1,n_2,\ldots,n_M),\quad 0\leq n_j\leq\statves_j,\ j=1,\ldots,M\Bigr\}.
$$
Here, $n_j$, $j=1,\ldots,M$ are the shells occupation numbers. For every configuration ${\bf n}\in\config$, define the energy of the zeroth-order approximation (for independent electrons in the effective central field)
\begin{equation}\label{ZeroEnergy}
E^{(0)}({\bf n})=\sum_{j=1}^Mn_j\eps_j
\end{equation}
and average configuration energy including the Coulomb interaction of the bound electrons unaccounted for in the zeroth-order approximation is
\begin{equation}\label{FirstEnergy}
E({\bf n})=\sum_{j=1}^Mn_jq_j+\frac 12\sum_{1\leq j,k\leq M}n_j(n_k-\delta_{kj})\theta_{jk}
\end{equation}
where 
$$
q_j=\eps_j+\int_0^{\infty}R_{jj}(x)\left(V(x)-\frac{Ze^2}{x}\right)dx
,\quad \theta_{jk}=\frac{\statves_j}{\statves_j-\delta_{jk}}\left(F_{jk}^{(0)}-\frac 14\sum_{s=0}^{\infty}
\mathfrak{R}_{s\kap_j\kap_k}G_{jk}^{(s)}\right),
$$
as usual (e.~g.~\cite{Nikiforov2005}, \cite{BarShalom1989}, \cite{Ovechkin2014}), are expressed through the Slater integrals
$$
F_{jk}^{(0)}=e^2\int_0^{\infty}\int_0^{\infty}\frac 1{x_>}R_{jj}(x_1)R_{kk}(x_2)dx_1dx_2,\quad
G_{jk}^{(s)}=e^2\int_0^{\infty}\int_0^{\infty}\frac{x_<^s}{x_>^{s+1}}R_{jk}(x_1)R_{jk}(x_2)dx_1dx_2,
$$
$$
R_{jk}(x)=f_j(x)f_k(x)+g_j(x)g_k(x),\ x_>=\max\{x_1,x_2\},\ x_<=\min\{x_1,x_2\};
$$
the angular factor
$$
\mathfrak{R}_{s\kap_j\kap_k}=\frac{(\kap_j+\kap_k-s)(\kap_j+\kap_k+s+1)}{\kap_j\kap_k}\cdot\wigner{s}{\ell(\kap_j)}{\ell(\kap_k)}^2,
$$
$$
\ell(\kap)=|\kap|-\heaviside(-\kap)
$$
and
$$
\heaviside(u)=\left[\begin{array}{ll}1,&u\geq 0\\0,&u<0\end{array}\right.,\,u\in\RR;\quad
\delta_{kl}=\left[\begin{array}{ll}1,&k=l\\0,&k\neq l\end{array}\right.,\,k,l\in\ZZ
$$
represent the Heaviside step function and the Kronecker delta, respectively. Note that the expression for $q_i$ can be modified to take into account the interaction with free electrons more accurately (see, e.~g.~\cite{Ovechkin2014}).

Consider a locally thermodynamically equilibrated plasma. Let $T$ be its temperature, $\beta = (k_{\rm B}T)^{-1}$; $\mu$ be the chemical potential, $\eta =\beta\mu$. 
The distribution over the set of configurations $\config$ is assumed to be Gibbsian, corresponding to the energies \eqref{ZeroEnergy}
$$
P_{\bf n}=\frac{1}{\Xi(\beta,\eta)}\statves({\bf n})\exp{-\beta E^{(0)}({\bf n})+\eta\sum_{j=1}^Mn_j}
\quad \statves({\bf n})=\prod_{j=1}^M\binom{\statves_j}{n_j},
$$
$$
\Xi(\beta,\eta)=\sum_{{\bf n}\in\config}\statves({\bf n})\exp{-\beta E^{(0)}({\bf n})+\eta\sum_{j=1}^Mn_j}
$$
and that, as it is well-known (e.~g.~\cite{Carson1968}), results in the binomial distribution 
\begin{equation}\label{GibbsBinom}
\Xi(\beta,\eta)=\prod_{j=1}^M(1+\exp{-\beta\eps_j+\eta})^{\statves_j},\quad
P_{\bf n}=\prod_{j=1}^M\binom{\statves_j}{n_j}p_j^{n_j}(1-p_j)^{\statves_j-n_j}
\end{equation}
where $p_j=(1+\exp{\beta\eps_j-\eta})^{-1}$ represents the occupation fractions of the shells. 

We will describe the bound-free opacity by means of a cross-section per atomic cell. For this cross-section, we start from the expression (see \cite{Peyrusse1999}, \cite{Bauche2015})
\begin{equation}\label{SigmaBFInitial}
\sigma_{\rm bf}(\omega)=\frac{2\pi^2\alpha a_0e^2}{\hbar\omega}\sum_{{\bf n}\in\config}P_{\bf n}\sum_{i=1}^M
n_i\sum_{\kap\in\ZZ\setminus\{0\}}\left(\heaviside(\eps)\mathfrak{f}_{i,\eps\kap}\right)\biggr|_{\eps=\hbar\omega-I_i({\bf n})}.
\end{equation}
The minor difference of \eqref{SigmaBFInitial} from the one in \cite{Nikiforov2005} is the absence of multiplier $(1+\exp{\beta\eps-\eta})^{-1}$. 
The bound--free analogue of the one-electron transition oscillator strength (oscillator strength density) in \eqref{SigmaBFInitial} is calculated according to the expression (e.~g.~\cite{Nikiforov2005})
\begin{multline}\label{OscillatorStrengthDensity}
\mathfrak{f}_{a,\eps\kap}=\frac{2mc^2}3\left(
\frac{\delta_{\kap_a+\kap,0}}{4\kap_a^2-1}+\frac{\kap}{\kap_a+\kap}\delta_{|\kap_a-\kap|,1}\right)\times\\\times\left(
(\kap_a-\kap-1)\int_0^{\infty}f_a(x)g_{\eps\kap}(x)dx+(\kap_a-\kap+1)\int_0^{\infty}f_{\eps\kap}(x)g_a(x)dx\right)^2
\end{multline}
for $a=1,\ldots,M$, $\eps>0$, $\kap\in\ZZ\setminus\{0\}$. It is important to note that the oscillator strength spectral density given by \eqref{OscillatorStrengthDensity} does not depend on configuration.

The configurations with $n_i=0$ give no contribution to the inner sum in \eqref{SigmaBFInitial}.
For the rest of configurations we take the ionization energy $I_{i}({\bf n})$ in \eqref{SigmaBFInitial} in the form 
\begin{equation*}
I_{i}({\bf n})=E({\bf n}-{\bf e}_i)-E({\bf n})=
-q_i+\sum_{j=1}^M(n_j-\delta_{ij})(-\theta_{ij}).
\end{equation*}
Here, ${\bf n}-{\bf e}_i$ is the configuration obtained from ${\bf n}$ by removing  one electron from shell $i$. We use the notation ${\bf e}_j$ for the vector in $\RR^M$ with coordinates $\delta_{sj}$, $s=1,\ldots,M$.
\section{The Convolution Representation of Bound--Free Opacity Cross-Section}
Let us introduce the notation 
\begin{equation}\label{DefQ}
Q_i(\eps)=\heaviside(\eps)\sum_{\kap\in\ZZ\setminus\{0\}}\mathfrak{f}_{i,\eps\kap}.
\end{equation}
The sum on the right-hand side of \eqref{DefQ} is finite. The function $Q_i(\eps)$ is zero for $\eps<0$ and decreases according to a power law when $\eps\to\infty$. The exponent of this power depends on $\kap_i$, but in any case it ensures the integrability of $Q_i$. Therefore, the Fourier transform is applicable to $Q_i$
$$
\widetilde Q_i(t)=\int_0^{\infty}\exp{\ii t\eps/\hbar}Q_i(\eps)d\eps,\quad
Q_i(\eps)=\frac{1}{2\pi\hbar}\int_{\RR}\exp{-\ii t\eps/\hbar}\widetilde Q_i(t)dt.
$$
Similarly to Fourier transformation method of DCA (see \cite{Arapova2025}), we change the order of summations in \eqref{SigmaBFInitial} and use simple properties of binomial coefficients to obtain
\begin{multline}\label{SigmaBF1}
\sigma_{\rm bf}(\omega)=\frac{2\pi^2\alpha a_0e^2}{\hbar\omega}\sum_{i=1}^M
\sum_{{\bf n}\in\config}P_{\bf n}
n_iQ_i(\eps)\Bigr|_{\eps=\hbar\omega-I_i({\bf n})}=\\=
\frac{\pi\alpha a_0e^2}{\hbar^2\omega}\sum_{i=1}^M\int_{\RR}\exp{-\ii t\omega}\widetilde Q_i(t)
\sum_{{\bf n}\in\config}P_{\bf n}
n_i\exp{\ii tI_i({\bf n})/\hbar}dt=\\=
\frac{\pi\alpha a_0e^2}{\hbar^2\omega}\sum_{i=1}^M\statves_ip_i\int_{\RR}\exp{-\ii t(\omega-\omega_i)}\widetilde Q_i(t)\Phi_i(t)dt
\end{multline}
where 
$$
\omega_{i}=u_i+\sum_{j=1}^M(\statves_j-\delta_{ij})p_jw_{ij},
\quad u_i=-q_i/\hbar,\quad w_{ij}=-\theta_{ij}/\hbar
$$
$$
\Phi_{i}(t)=\prod_{j=1}^M\left((1-p_j)\exp{-\ii t p_jw_{ij}}+p_j\exp{\ii t(1-p_j)w_{ij}}\right)^{\statves_j-\delta_{ij}}.
$$
Unlike the case of the bound--bound opacity (see \cite{Arapova2025}), the computation of the expression \eqref{SigmaBF1} using the numerical Fourier transform is inconvenient due to the behavior of $Q_i$. To fix this, we notice that $\Phi_i$ is the characteristic function (e.~g.~\cite{Shiryaev1996}) of the centered linear combination
\begin{equation}\label{LinearCombinationRVs2}
X_i=\sum_{j=1}^Mw_{ij}(\xi_{ij}-\me\xi_{ij})
\end{equation}
of independent random variables $\xi_{ij}\sim{\rm Binomial}(\statves_j-\delta_{ij},p_j)$, $j=1,\ldots,M$. Recall that $\xi\sim{\rm Binomial}(m,p)$, $m\in\ZZ_+$, $p\in[0,1]$ means that
$$
\prob(\xi=k)=\binom{m}{k}p^k(1-p)^{m-k},k=0,\ldots,m.
$$
Obviously, $\me X_i=0$ and 
\begin{equation}\label{VarianceXi}
D_i=\me X_i^2=\var X_i=\sum_{j=1}^M(\statves_j-\delta_{ij})p_j(1-p_j)w_{ij}^2<\infty.
\end{equation}
We use standard notations $\prob$, $\me$ and $\var$ for the probability of an event, the mathematical expectation of a random variable and its variance. Further, let 
\begin{equation}\label{DefFi}
F_i(y)=\prob(X_i\leq y),\quad 
\Phi_i(t)=\int_{\RR}\exp{\ii t y}dF_i(y)
\end{equation}
be correspondingly the distribution function and the characteristic function of $X_i$.
Using $F_i$, we can transform the right-hand side of \eqref{SigmaBF1} to the form
\begin{equation}\label{SigmaBFFinal}
\sigma_{\rm bf}(\omega)=\sum_{i=1}^M\statves_ip_i\sigma_{{\rm bf},i}(\omega),\quad 
\sigma_{{\rm bf},i}(\omega)=
\frac{2\pi^2\alpha a_0e^2}{\hbar\omega}\left.\left(\int_{-\infty}^{\eps/\hbar}Q_i(\eps-\hbar y)dF_i(y)\right)\right|_{\eps=\hbar(\omega-\omega_i)},
\end{equation}
we introduced the cross-section $\sigma_{{\rm bf},i}$ of $i$-shell photoionization. 
Applying Chebyshev's inequality (e.~g.~\cite{Shiryaev1996}), it is easy to see that the distribution of $X_i$ is practically concentrated on the interval $[-LD_i^{1/2},LD_i^{1/2}]$ for large enough $L$. To calculate the convolution integral in the right-hand side of \eqref{SigmaBFFinal}, we will choose $N$ and use piecewise linear interpolation of $F_i(y)$ over the points
$$
y_l=2LD_i^{1/2}l/N,\ l=-N/2,\ldots,N/2-1
$$
to obtain an approximate expression for the integral in \eqref{SigmaBFFinal}
\begin{equation}\label{ConvolutionCalculation}
\int_{-\infty}^{\eps/\hbar}Q_i(\eps-\hbar y)dF_i(y)\approx
\sum_{l=-N/2+1}^{N/2-1}\frac{F_i(y_l)-F_i(y_{l-1})}{y_l-y_{l-1}}\frac 1{\hbar}\int_{\eps-\hbar y_l}^{\eps-\hbar y_{l-1}}Q_i(\epsilon)d\epsilon.
\end{equation}
The integrals of $Q_i$ are calculated using piecewise log-linear interpolation over the grid values of $Q_i$ suitable for power-law asymptotics. In the calculations below, we assumed $L=5$, $N=1024$.

To determine the distribution function of $X_i$ from its characteristic function $\Phi_i$, we follow \cite{Bohmann1975} in using the square integrability of $X_i$ and the identity $\Phi_i(-t)=\Phi_i^*(t)$ to express
\begin{multline}\label{DistributionFunc}
F_i(y)=\frac 12\left(1+\erf\left(\frac{y}{\sqrt{2D_i}}\right)\right)+\frac{\ii}{2\pi}\int_{\RR}\frac{\Phi_i(t)-\exp{-D_it^2/2}}{t}\exp{-\ii ty}dt=\\
=\frac 12\left(1+\erf\left(\frac{y}{\sqrt{2D_i}}\right)\right)+2{\rm Re}\int_0^{\infty}\frac{\ii(\Phi_i(t)-\exp{-D_it^2/2})}{2\pi t}\exp{-\ii ty}dt.
\end{multline}
The integrands in \eqref{DistributionFunc} have no singularity at the origin because $(\Phi_i(t)-\exp{-D_it^2/2})/t^2\to 0$ when $t\to 0$. For the selected values $y_l$ we get the following approximation  for the last integral in \eqref{DistributionFunc}
$$
2{\rm Re}\int_0^{\infty}\frac{\ii(\Phi_i(t)-\exp{-D_it^2/2})}{2\pi t}\exp{-\ii ty_l}dt\approx\frac{1}{L\sqrt{D_i}}{\rm Re}\sum_{k=1}^{N-1}\frac{\ii(\Phi_i(t_k)-\exp{-D_it_k^2/2})}{t_k}\exp{-\ii t_ky_l},\quad t_k=\frac{\pi k}{L\sqrt{D_i}}.
$$
The produced sums can be evaluated using fast Fourier transform. 

Leaving only the first term on the right-hand side of \eqref{DistributionFunc} corresponds to the Gaussian approximation of the distribution of $X_i$. This approximation gives the following expressions for the required convolution integrals 
\begin{equation}\label{GaussBF}
\sigma_{{\rm bf},i}(\omega)\approx\frac{2\pi^2\alpha a_0e^2}{\hbar\omega(2\pi D_i)^{1/2}}
\int_{-\infty}^{\eps/\hbar}Q_i(\eps-\hbar y)
\exp{-y^2/(2D_i)}dy
\end{equation}
that is similar to the one recieved in \cite{Dragalov1989}, \cite{Nikiforov2005} (with slight differences in the expressions for $D_i$). If $\sqrt{D_i}$ is significantly less than the width of the spectral interval, where the cross-section has to be determined, one can also use the simplification $F_i(y)\approx\heaviside(y)$ that causes the zero threshold width approximation
\begin{equation}\label{HeavisideBF}
\sigma_{{\rm bf},i}(\omega)\approx\frac{2\pi^2\alpha a_0e^2}{\hbar\omega} Q_i(\hbar(\omega-\omega_i)).
\end{equation}

\section{Calculation results and discussion}
To illustrate the described method, we consider the calculation of the bound-free opacity cross-section for iron plasma under several conditions. For the calculations, we used the average characteristics of the atom obtained within the Dirac --- Hartree --- Fock --- Slater approximation according to \cite{Nikiforov2005}. To find the bound states, we applied the phase method \cite{Vronskiy2018:2}.
The states of free electrons were normalized according to the condition \eqref{ContinuousNormalize} and
\begin{equation}\label{VZero}
V(x)=0\mbox{ when } x\geq x_{\rm cell}=\left(\frac{3A}{4\pi N_{\rm Av}\rho}\right)^{1/3},
\end{equation}
we used notations $\rho$ for plasma density, $A$ for the atomic weight and $N_{\text{Av}}=6.022\cdot 10^{23}$ for Avogadro's number. Condition \eqref{VZero} leads to the representation of the Dirac equation solution as a combination of spherical Bessel functions (e.~g.~\cite{Starrett2015}, \cite{Trzhaskovskaya2018}). 

\begin{figure}[ht!]
\centerline{\includegraphics[width=0.7\textwidth]{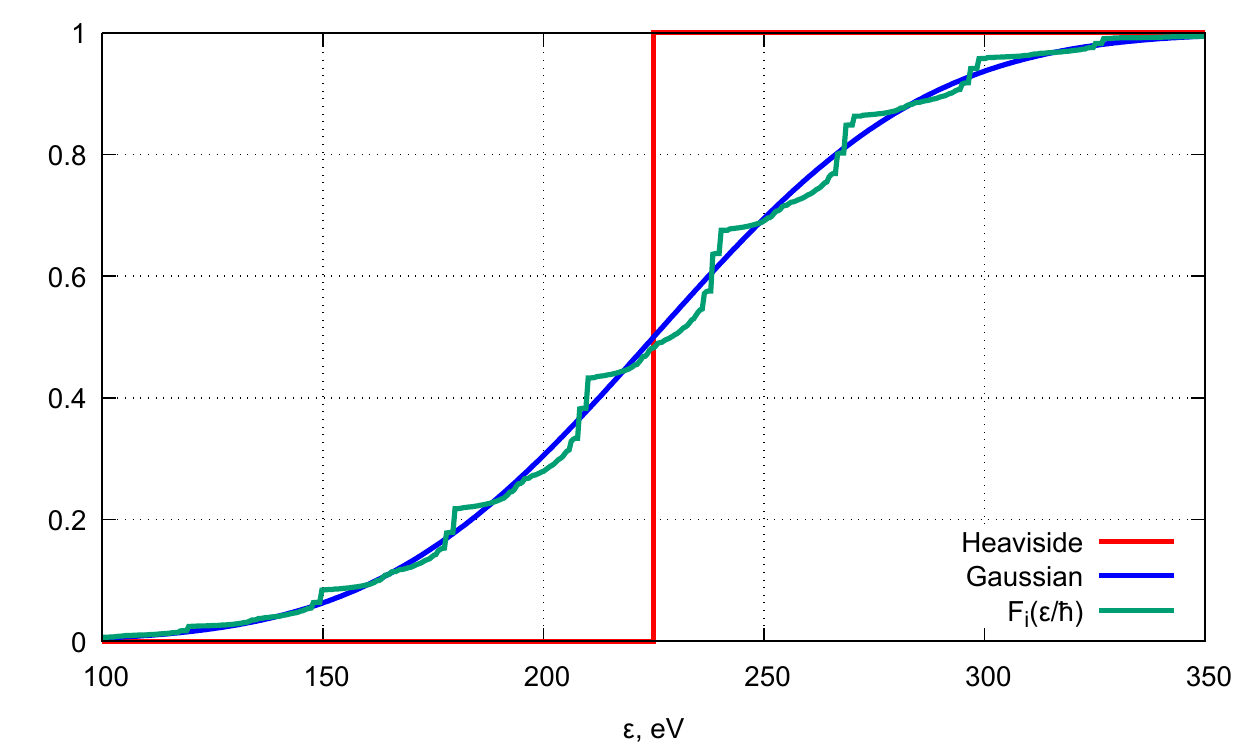}}
\caption{The distribution functions of $3s$-shell ionization energy for iron with $T=20\ev$, $\rho=0.01\gccm$}\label{Distr3s}
\end{figure}

\begin{figure}[ht!]
\centerline{\includegraphics[width=0.7\textwidth]{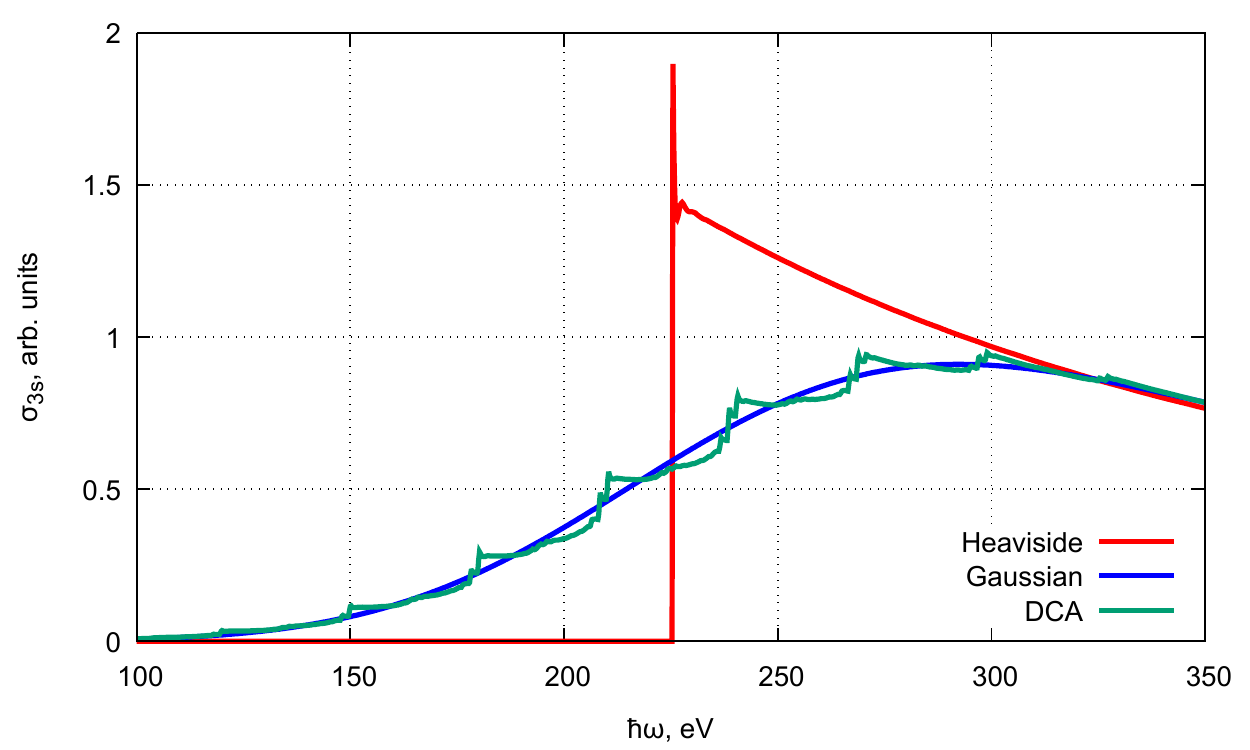}}
\caption{The $3s$-shell photoionization cross-section for iron plasma, $T=20\ev$, $\rho=0.01\gccm$}\label{Sigma3s}
\end{figure}

\begin{figure}[ht!]
\centerline{\includegraphics[width=0.7\textwidth]{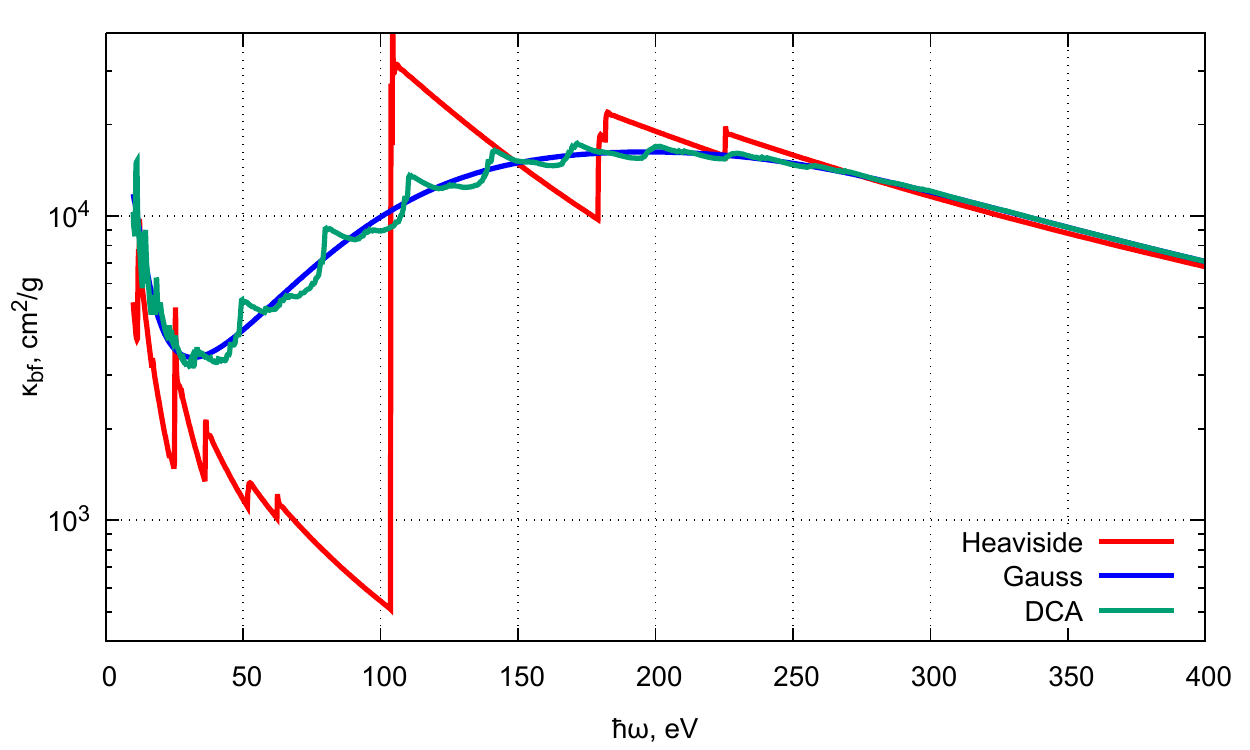}}
\caption{The bound-free opacity coefficient for iron plasma, $T=20\ev$, $\rho=0.01\gccm$}\label{Fe20}
\end{figure}

\begin{figure}[ht!]
\centerline{\includegraphics[width=0.7\textwidth]{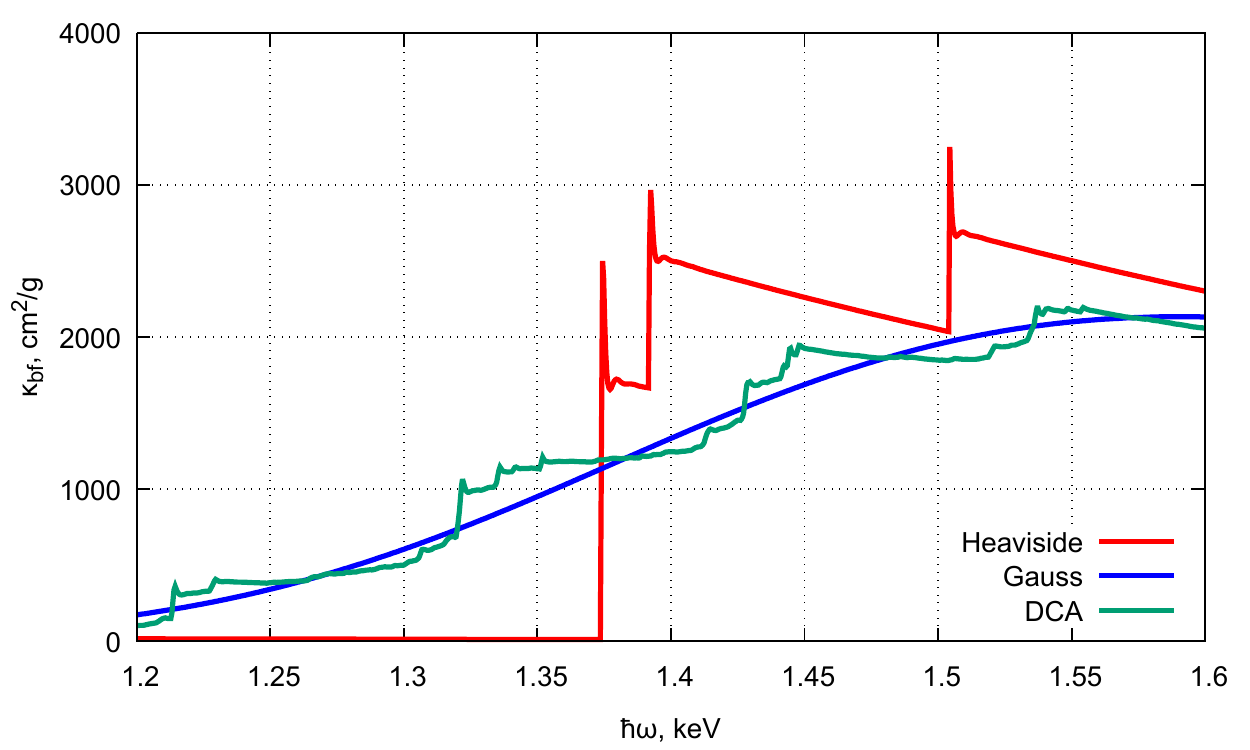}}
\caption{The bound-free opacity coefficient for iron plasma, $T=150\ev$, $\rho=0.02\gccm$}\label{Fe150}
\end{figure}

In Fig.~\ref{Distr3s} the distribution function $F_i(\eps/\hbar)$ \eqref{DefFi} of the $3s$-shell ionization energy for iron plasma with temperature $20\ev$ and density $0.01\gccm$ is plotted together with its Gaussian approximation (the first term on the right-hand side \eqref{DistributionFunc}) and the degenerate distribution $\heaviside(\eps-\hbar\omega_{3s})$. In Fig.~\ref{Sigma3s} the photoionization cross-sections of the $3s$-shell corresponding to these distributions are shown: obtained by the method proposed in this paper \eqref{SigmaBFFinal}, \eqref{ConvolutionCalculation}, using the Gaussian approximation \eqref{GaussBF} and without taking into account the occupation number fluctuations \eqref{HeavisideBF}. In Fig.~\ref{Fe20} the total bound-free opacity coefficient $\kappa_{\rm bf}(\omega)=N_{\text{Av}}\sigma_{\rm bf}(\omega)/A$, obtained by the same three methods are presented. In Fig.~\ref{Fe150} the similar values of $\kappa_{\rm bf}(\omega)$ for iron plasma at $T=150\ev$ and $\rho=0.02\gccm$ are presented.

The results make it evident that the Gaussian approximation satisfactorily describes the spectrum-averaged behavior of the bound-free opacity coefficient near the threshold for the examples considered. In some areas of the spectrum the difference is quite significant. Local maxima obtained by the proposed method of DCA correspond to steeper areas of the distribution functions $F_i$, that are mainly determined by transitions from ions of a certain charge.

\section{Conclusions}
In this work, we propose the method of the Detailed Configuration Accounting for the bound-free opacity calculations. This method extends the Fourier transform method of DCA for calculating the bound-bound opacity, which, in its turn, is the adaptation of the Hazak-Kurzweil method of configurationally resolved STA. The final formulas for the bound-free opacity cross-sections are given in the form of convolution of spectral oscillator strength density with the distribution function of threshold energy. Simplifying approximations are considered as well.
To illustrate the method, we used an average atom model with Dirac one-electron states. At the same time, the method is also applicable for another ways of describing one-electron states: Schrödinger orbitals, parametric models, etc. The basic assumption for the proposed method is the factorization of the probabilities of configurations (i.e., the statistical independence of the shell occupation numbers). We have neglected the detailed structure of the threshold for the bound-free transition from a particular configuration: the corresponding effect is less significant than in the bound-bound case.

\section*{Acknowledgments}
The authors are grateful to A.~A.~Ovechkin and E.~S.~Tsoy for useful discussions.

\end{document}